# Pseudomagnetic Fields in a Locally Strained Graphene Drumhead

Shuze Zhu,[1] Yinjun Huang,[1] Nikolai K. Klimov,[2,3] David B. Newell,[4] Nikolai B. Zhitenev,[3] Joseph A. Stroscio,[3] Santiago D. Solares,[1,2,3] Teng Li[1,2*]

[1]Department of Mechanical Engineering, University of Maryland, College Park, MD 20742, USA

[2]Maryland NanoCenter, University of Maryland, College Park, MD 20742, USA

[3]Center for Nanoscale Science and Technology, NIST, Gaithersburg, MD 20899, USA

[4] Physical Measurement Laboratory, NIST, Gaithersburg, MD 20899, USA

**Abstract:** Recent experiments reveal that a scanning tunneling microscopy (STM) probe tip can generate a highly localized strain field in a graphene drumhead, which in turn leads to pseudomagnetic fields in the graphene that can spatially confine graphene charge carriers in a way similar to a lithographically defined quantum dot (QD). While these experimental findings are intriguing, their further implementation in nanoelectronic devices hinges upon the knowledge of key underpinning parameters, which still remain elusive. In this paper, we first summarize the experimental measurements of the deformation of graphene membranes due to interactions with the STM probe tip and a back gate electrode. We then carry out systematic coarse grained, (CG), simulations to offer a mechanistic interpretation of STM tip-induced straining of the graphene drumhead. Our findings reveal the effect of (i) the position of the STM probe tip relative to the graphene drumhead center, (ii) the sizes of both the STM probe tip and graphene drumhead, as well as (iii) the applied back-gate voltage, on the induced strain field and corresponding pseudomagnetic field. These results can offer quantitative guidance for future design and implementation of reversible and on-demand formation of graphene QDs in nanoelectronics.

---

[*] Corresponding author: Email: lit@umd.edu

# I. Introduction

Graphene, with its capability to accommodate a wealth of remarkable properties, has recently received extensive attention [1-3]. Many of its properties are strongly tied to the symmetry of its hexagonal crystal lattice [4]. Both in-plane and out-of-plane deformations lead to the change or breaking of its crystal lattice symmetry, and thus can also alter its properties [5]. This opens up fertile opportunities to leverage strain engineering, to tailor its properties, or even render its unconventional but desirable properties, in order to enable graphene-based electronic devices [6]. For example, strain-induced pseudomagnetic fields in graphene have been explored as a potential approach to engineering its electronic states. Pseudomagnetic fields as strong as 300 T have been shown to exist in highly strained graphene nanobubbles, which are formed naturally during cooling down of graphene grown by chemical vapor deposition on a platinum substrate to cryogenic temperatures, due to a relatively large difference between the thermal expansion coefficients of graphene and platinum [7]. Recent pioneering scanning tunneling microscopy/spectroscopy (STM/STS) experimental studies of suspended graphene drumheads and related molecular dynamics simulations revealed that an STM probe tip can interact with a suspended graphene drumhead and generate a highly localized radially symmetric strain field in graphene right under the STM probe tip [2]. This strain field, in turn, leads to a pseudomagnetic field of three-fold symmetry in graphene [8]. The STS experimental study reveals that such pseudomagnetic fields induced in a freestanding graphene membrane give rise to a repeating series of a four-multiplet peak structure in the local density of states, which was interpreted as confinement of graphene charge carriers in a quantum dot (QD) [2]. The STS experimental data also suggest that although the tip-induced QD is quite leaky, an external applied magnetic field can further improve the overall spatial confinement of graphene charge carriers. In addition, the simulations show that such a QD formed right under the STM probe tip follows the tip trajectory as it rasters across the graphene drumhead, while the resulting pseudomagnetic field and QD size remain nearly unchanged. Given the large elastic deformability of graphene and the retractability of the STM probe tip, the above results suggest a facile approach toward reversible and on-demand formation of graphene QDs, a desirable feature for graphene-based nanoelectronics [9-13]. Promising potential aside, further implementation of such a feature hinges upon quantitative understanding of several issues that still remain unexplored. For example, how do the intensity of the resulting pseudomagnetic field and the QD size depend on the characteristic size of the graphene drumhead as well as on the size of the STM probe tip? What are the effects of back-gate voltage and probe tip position on the resulting pseudomagnetic field and QD? Answers to these questions are crucial to achieve spatially confined QDs in graphene with tunable intensity and size.

To address the aforementioned unanswered questions, this paper presents a systematic study using a coarse-grained (CG) modeling scheme to offer a mechanistic interpretation of STM tip-induced straining of a graphene drumhead. Our investigations reveal how the position of the STM probe tip relative to the graphene drumhead center, the sizes of both the STM probe tip and graphene drumhead, as well as the applied back-gate voltage, affect the induced strain field and corresponding pseudomagnetic field. The predictions of our simulations are validated by agreement with recent experimental measurements [2], and thus the results from our parametric modeling study can serve as guidelines for future design and implementation of reversible and on-demand formation of graphene QDs in nanoelectronics.

## II. Experimental Results

To facilitate the description of the simulation model, it is necessary to summarize the graphene device structure and STM results used in the aforementioned recent experiments revealing QD formation in suspended graphene drumheads [2]. One of the back-gated graphene devices used in the experiments of Ref. [2] is shown on Fig. 1(a) and (b). The device consists of a single layer of graphene placed over an array of pits etched in a $SiO_2$/Si substrate. The array of pits, with pit diameter and depth of 1.1 μm and 100 nm, respectively, was fabricated via photolithography followed by plasma etching of $SiO_2$ [Fig. 1(b)]. Graphene flakes were placed onto the pre-patterned $SiO_2$/Si substrate via mechanical exfoliation of natural graphite. An electrode for biasing the graphene was fabricated via e-beam assisted evaporation of a metal electrode through a stencil mask in order to minimize contamination of graphene during the device fabrication process. *P*-doped Si in a $SiO_2$/Si substrate was used as a back gate electrode to tune the type and the number of charge carriers in the graphene sheet.

The measurements were done in a home built ultrahigh vacuum (UHV) cryogenic STM system operating at 4K [14]. The alignment of the iridium STM probe tip relative to the device was done *in situ* in UHV via an external optical telescope. In the experiment, stable STM measurements on the suspended graphene were achieved by carefully approaching the graphene membrane with very slow scanning speeds. Both the van der Waals (vdW) force from the STM probe tip and an electrostatic force from the applied back-gate voltage induced substantial mechanical deformations in the suspended graphene membrane, as shown in Fig. 2. The deformation could be tuned to be either outward from the surface or inward into the pit under the suspended graphene depending on the strength of the back gate potential, $V_{GATE}$. The graphene deformation in Fig. 2 is caused by the STM probe tip pulling up on the membrane and the electric field from the back gate electrode pulling it down. As seen in Fig. 2, the deformation starts to curve downward into the pit when the force from the back gate electrode overtakes the force from the tip at a back gate potential of ≈ 50 V. In contrast, we did not observe a substantial effect on the deformation from the tip electric field. Figure 3(a) shows the membrane profile for a series of tip potentials at a fixed back gate potential. While some trend of larger deformation with increasing tip potential is seen in Fig. 3(a), the effect is much smaller than that from the back gate potential, and almost within the noise or variations obtained when scanning over the membrane. These results can be understood based on consideration of the size of the probe tip relative to membrane and how the membrane is shaped by the tip forces. The typical global diameter of our STM probe tips is on the order of 110 nm [Fig. 3(b)]. As the simulations below will show, the graphene deformations are highly localized in the region near the STM probe tip resulting in a tent-like apex, as shown schematically in Fig. 1(c). The lack of an effect from the tip potential variation in the STM profiles in Fig. 3(a) indicates that the vdW forces from the probe tip dominate the force from the tip potential. However, the electric force from the probe tip may be somewhat masked from how the STM servo is adjusting the probe tip height to keep a constant tunneling gap. As both the vdW force and probe tip electric field pull on the graphene membrane, the STM feedback loop pulls the tip back to the point of maintaining a tunneling gap set by the tunneling current setpoint, *i.e.*, just before it would release the graphene membrane. Therefore the additional and localized force from the tip potential may not significantly alter the graphene profile shape, which is dominated by the vdW force. It is important to note that the profiles observed in Fig. 2 are a result of dragging the top of the cusp of the pulled-up membrane as the tip is rastered, *i.e.*, the profiles in Fig. 2 are not the static shape of the graphene membrane.

We investigated the change in graphene electronic structure due to the graphene membrane deformation using scanning tunneling spectroscopy gate mapping [2]. In these measurements the differential conductivity is measured as a function of sample bias to reveal the graphene density of states. Measurements in magnetic fields reveal a series of Landau levels corresponding to the quantization of graphene charge carriers with positions that are varied with carrier density (gate voltage) [Fig. 4(a)][15].

As the tip approaches the membrane edge, the membrane can be delaminated while the graphene is still supported by the substrate within about 50 nm of the membrane edge. This can be seen in the membrane profiles in Fig. 3(a), and also in gate maps in this vicinity in Fig. 4(b). Once the tip is on the suspended membrane, the tunneling spectra in the gate map change from those of the graphene on the supported areas, as seen in Fig. 4(c). This indicates that the induced strain in the suspended graphene drumhead originating from the tip-membrane interactions dramatically alters the electronic spectrum of graphene charge carriers, compared to the electronic spectrum of graphene directly supported by a substrate, as seen by comparing Fig. 4(a) with Fig. 4(c).

The spectral signatures in the gate maps on the suspended graphene [Fig. 4(c)] resemble those from QDs [15, 16]. The spectral features consist of four-fold resonances with a positive gate voltage slope. A single spectrum from the gate map in Fig. 4(c), at a gate voltage of 5 V, is plotted in Fig. 4(d) showing multiple groups of four peaked resonances. Each resonance peak corresponds to the opening of a new transport channel at the Fermi level associated with a single electron addition to the QD. The lines in the gate map are thus the lines of constant chemical potential and are tilted because the QD levels are controlled by a linear combination of the bias voltage and gate voltage. We can extract the charging energies of the QDs from an analysis of the resonance peak positions. The peak positions of these resonances are plotted in Fig. 4(e), and the QD addition energies [Fig. 4(f)] are given by the differences in the resonance peak energies scaled by the bias lever arm, $E = \alpha V_\text{B}$, where $\alpha = \frac{C_\text{B}}{C_\text{T}} \approx 0.4$, $C_\text{B}$ is the graphene layer to graphene QD capacitance, and $C_\text{T}$ is the total capacitance of the QD [2]. The addition energies in Fig. 4(f) follow the classic spectrum of a QD [17] with energies given by $\frac{e^2}{C_\text{T}} + \Delta E_{N+1}$, where $e$ is the electron charge and $e^2/C_\text{T}$ is the charging energy of the fourfold degenerate levels in a graphene QD. The energy $\Delta E_{N+1} = E_{N+1} - E_N$ separating each group of four resonances corresponds to the energy required to reach the next QD level. The range of addition energies in the gate map in Fig. 4(c) correspond to QDs with diameters ranging from $\approx 34$ nm to $\approx 53$ nm [2].

## III. Simulation Model Setup

Given the relatively large sizes of both the suspended graphene drumhead ($\approx 1.1$ μm diameter) and the STM probe tip ($\approx 110$ nm radius) used in the experiments, fully atomistic simulations of the STM tip-induced deformations in the graphene drumhead are prohibited due to the limit of computation capacity. To model the related tip induced deformations in graphene, we adopt a scalable bottom-up CG simulation scheme [18], which is recapped below. Figure 5(a) shows the carbon atom representations of CG beads of the first three orders. Note that a CG scheme of $N^{\text{th}}$ order allows for a reduction of computation model size by $4^{N+1}$ times. For example, one $3^{\text{rd}}$ order CG bead represents 256 carbon atoms, which significantly reduces the computational expense. One unique feature of this CG scheme is its self-similarity in lattice structure. That is, the hexagonal lattice structure of graphene, the origin of many exceptional graphene properties, is maintained in the CG lattice structure of any order [Fig. 5(b)]. The mass of the CG bead and the bond length between two CG beads of any order can be deduced recursively. Due to the nature of the vdW interaction, only a small portion of the STM probe tip interacts with the graphene drumhead. Without loss of generality, in the simulation model we coarse-grain the STM probe tip as a spherical bead, which approximates the real shape of the probe tip [Fig. 3(b)], and since the global shape of the real STM probe tip in experiments plays a minor role in determining the deformation in the graphene drumhead. As the vdW interaction decays significantly beyond 1 nm, the interaction from the tip will mainly result from the portion of the tip that is in very close proximity to the membrane. Therefore the shape of the tip beyond a few nm away from the membrane is not very significant. To better simulate the STM experiment, the diameter of the CG spherical bead is taken to be equal to the curvature of the real STM probe tip. The left panel in Fig. 5(c) describes a typical CG computational model in the present study. The tip is positioned slightly above the membrane at the desired horizontal position, relative to the center of the graphene drumhead and is held at a fixed position while the membrane relaxes with its edge atoms fixed. The choice of a proper order of CG scheme is tip-size and graphene-size dependent. A proper order of CG scheme is selected so that, on one hand, the CG graphene bead lattice is small enough to capture the fine features of the localized deformations in the graphene drumhead, and on the other hand, large enough to reasonably reduce the computational cost.

The energy expression for the above CG scheme includes bonded energy terms and non-bonded energy terms. The bonded terms consist of two-body bond energy and three-body angle energy as

$$U_{bonding}(r_{ij},\theta_{ijk}) = \sum \frac{1}{2} K_b (r_{ij} - r_0)^2 + \sum \frac{1}{2} K_\theta (cos\theta_{ijk} - cos\theta_0)^2, \qquad (1)$$

where $K_b$ and $K_\theta$ are the bond force constant and angle force constant, respectively, $r_{ij}$ is the distance between the i$^{th}$ and j$^{th}$ CG beads while $r_0$ is its corresponding equilibrium distance; $\theta_{ijk}$

is the angle formed between the i-j bond and j-k bond and $\theta_0$ is the corresponding equilibrium angle. The STM tip-induced deformations in the graphene drumhead are largely due to stretching and bending, while twisting of the graphene lattice is expected to be of secondary significance. Therefore, energy terms for the torsion are not considered in the CG scheme to reduce the computation expense. The bond force constant is determined by equating the potential energies of an atomistic graphene structure and its CG counterpart under equal biaxial stretching (no change in the three-body angle energy), while the angle force constant is determined by equating the potential energies of an atomistic graphene structure and its CG counterpart under equal biaxial bending (in a scenario where changes in the two-body bond energy are highly suppressed). It turns out that the bond force constant is the same for all CG levels and can be determined from the Morse potential for C-C bonds [19] to be 47.46 eV·Å$^{-2}$. The angle force constants are analytically derived to be 93.23 eV, 372.91 eV and 1491.66 eV for the 1$^{st}$, 2$^{nd}$ and 3$^{rd}$ order CG schemes, respectively. The non-bonded term includes the vdW interaction between the CG graphene beads and the CG STM tip bead, which is fitted using a shifted Lennard-Jones (LJ) potential as

$$E = 4\epsilon \left[ \left( \frac{\sigma}{r-\Delta} \right)^{12} - \left( \frac{\sigma}{r-\Delta} \right)^{6} \right], \tag{2}$$

with a variable $\Delta$ to accommodate a CG STM tip with a non-zero diameter. The parameters in this shifted LJ potential can be determined by equating its value to the total vdW interaction energy between an atomistic iridium spherical tip and a triangular atomistic graphene flake represented by CG graphene beads of a given order (e.g., as in Fig. 5(a)). The interatomic pair interaction between iridium and carbon atoms is modeled using the conventional LJ 6-12 potential, the parameters of which are determined through the customary Lorentz-Berthelot mixing rules, using the iridium-iridium and carbon-carbon parameters from the universal force field [20]. All CG simulations were implemented within the Large-scale Atomic/Molecular Massively Parallel Simulator (LAMMPS) [21].

The right panel of Fig. 5(c) compares CG simulation results with fully atomistic molecular dynamics simulation results of the STM tip-induced deformations in a graphene drumhead with a diameter of 50 nm. Here the STM probe tip is positioned above the center of the graphene drumhead and there is no electrostatic force from the back gate electrode. In the CG simulations, the STM probe tip is gradually retracted from the graphene drumhead until a critical point is reached, at which further retraction of the tip would result in the loss of tip-graphene contact. In each step of the CG simulations, the energy of the system is minimized by using the conjugate gradient algorithm until either the total energy change between successive iterations divided by the energy magnitude is less than or equal to $10^{-6}$, or the total force is less than $10^{-5}$ eV/ Å. As seen in Fig. 5(c) both the $1^{st}$ and $2^{nd}$ order CG simulation schemes agree quite well with fully atomistic simulations of a 50 nm graphene drumhead. This justifies the validity of our CG simulation scheme.

**IV. Results and Discussion**

*IV.1. STM tip-induced localized deformations of graphene drumheads and the associated pseudomagnetic field*

Following the CG modeling strategy described in Section III, we next apply the $3^{rd}$ order CG scheme to study the STM tip-induced localized deformation in the graphene drumhead used in the experiment as well as the resulting pseudomagnetic field. We first consider the case without an electrostatic force from the back gate. The diameters of the STM probe tip and the graphene drumhead are taken to be 110 nm and 1100 nm, respectively. Figure 6(a) shows the contour plot of the out-of-plane deflection of the graphene drumhead, when the STM probe tip is center-positioned at the critical height. The induced deformations have an axisymmetric tent-like shape about the center as the out-of-plane deflection gradually increases away from the edge and peaks in the immediate vicinity of the STM probe tip. Figures 6(b)-6(d) show the contour plots of the components of the Lagrange strain tensor in the deformed graphene drumhead, with a peak value of normal strain up to 1.4 % and peak shear strain up to 0.55 %. It has been shown that the lattice distortions due to strain in graphene introduce effective gauge fields in the Dirac Hamiltonian, which shift the $K$ and $K'$ points in the Brillouin zone in opposite directions, much like the effect of an applied perpendicular magnetic field [6]. The resulting gauge field, $A$, leads to a pseudomagnetic field, $B_{ps}$, which acts on the electrons and hole charge carriers in graphene. The strain-induced pseudogauge field $A_{ps}$ is calculated from the strain components, $u_{ij}$, as [6, 8, 22, 23]

$$A_{ps} = \frac{t\beta}{ev_F}(u_{xx} - u_{yy}, -2u_{xy}), \qquad (3)$$

where $\beta = 2.5$ is the dimensionless coupling constant, $t = 2.8$ eV is the hopping energy and $v_F = 1 \times 10^6\ m\ s^{-1}$ is the Fermi velocity. The pseudomagnetic field can be computed as $B_{ps} = \nabla \times A_{ps}$. The resultant pseudomagnetic field can be denoted by a vector, whose magnitude can be expressed as

$$B_{ps} = \frac{t\beta}{ev_F}\left(\frac{\partial(-2u_{xy})}{\partial x} - \frac{\partial(u_{xx}-u_{yy})}{\partial y}\right). \qquad (4)$$

Equation (4) shows that the intensity of the pseudomagnetic field scales with the magnitude of the strain gradient rather than with the strain. As a result, even though the STM tip-induced strain magnitude in the graphene drumhead is modest, the highly localized nature of the deformations dictates large strain gradients, which, in turn, leads to significant pseudomagnetic fields in the graphene drumhead. Furthermore, in a cylindrical coordinate system ($r$, $\theta$), for axially symmetrical situations, Eq. (4) can be transformed into

$$B_{ps} = \frac{t\beta}{ev_F} sin3\theta \left(-\frac{\partial(u_{rr}-u_{\theta\theta})}{\partial r} + \frac{2(u_{rr}-u_{\theta\theta})}{r}\right). \tag{5}$$

This suggests that a strain field in graphene having rotational symmetry can lead to a pseudomagnetic field with a threefold symmetry (as a result of the pre-factor $sin3\theta$), as opposed to a strain field with threefold symmetry leading to a uniform pseudomagnetic field [6]. Figures 6(e) and (f) present the pseudomagnetic field associated with the tip-induced localized strain field in the graphene drumhead, which assumes a clover-leaf threefold symmetry with alternating intensity peaks of about $\pm 10$ T. Our recent experiments further reveal that such pseudomagnetic fields in the graphene drumhead could directly affect the graphene electronic properties in a sense analogous to the charge carrier confining effect in a lithographically defined QD [2]. As shown in Fig. 6(e), the pseudomagnetic fields are mostly located at the drumhead's center within a region of ≈ 40 nm in diameter, which agrees well with the effective QD size (≈ 42 nm in diameter) estimated from the charging energies in the STS gate maps measured in the experiment [Fig. 4(f)] [2]. Such a good agreement on the estimated QD size from two distinct approaches serves as further evidence of the effectiveness and precision of the CG simulation scheme.

*IV.2. Dependence of the pseudomagnetic field on graphene membrane size and STM tip size*

We carry out a parametric study to investigate the effects of the STM tip size and the graphene drumhead diameter on the intensity of the strain-induced pseudomagnetic field and the associated QD size. It is shown that in all simulation cases, corresponding to the center-positioned STM probe tip, the interaction between the STM probe tip and the graphene drumhead leads to tent-like deformations of the graphene drumhead similar to that shown in Fig.

6(a). In general, for a given STM tip size, a larger graphene drumhead corresponds to an overall larger amplitude of the out-of-plane deformations. However, as shown in Eqs. (4) and (5), the resulting pseudomagnetic field depends much more strongly on the strain gradient than on the displacement and strain of the graphene drumhead. Therefore, it is the highly localized strain field in the graphene at the immediate vicinity of the STM tip that dictates the nature of the pseudomagnetic field and the associated QD formation.

Figure 7 summarizes the findings of the above parametric study. It shows the dependence of the magnitude of the induced pseudomagnetic field and the characteristic size of the formed QD on the diameters of the STM probe tip and the graphene drumhead. For a given STM tip size, there exists a threshold graphene diameter (about 400 nm, indicated by a dashed line in Figs. 7(a)-7(c)) that delineates two regimes for the above mentioned dependences. For graphene drumheads with diameters smaller than the threshold value, the intensity of strain-induced pseudomagnetic field decreases, while the corresponding QD size increases, as the graphene diameter increases. For such graphene drumheads, the smaller the STM tip size, the smaller the resulting QD, but the stronger the pseudomagnetic field. For example, for STM tip sizes of 45 nm, 75 nm and 110 nm and for a fixed graphene membrane diameter of 250 nm, the corresponding pseudomagnetic fields and QD sizes are 23 T, 16 T, 13 T, and 13 nm, 18 nm, 29 nm, respectively. Figure 7(d) further reveals that for the graphene drumhead of diameter comparable to those in the experiments (1100 nm), the intensity of pseudomagnetic field decreases while the QD size increases as the STM probe tip diameter increases. However we observe a saturation trend for both the intensity of pseudomagnetic field and QD size as the STM tip size continues increasing. The above dependence on the STM tip size can be understood as follows. The interaction force exerted by the STM probe tip on the graphene drumhead is to some extent analogous to the effect of a pole sticking out of a circular tent at its center (as is evident from the tent-like morphology of the graphene drumhead). A smaller STM tip acts like a sharper pole, which leads to a smaller size of the region with highly localized deformation in the tent (corresponding to the QD size in strained graphene), but a higher strain concentration in such a region (i.e., higher strain gradient) corresponds to a greater intensity of the pseudomagnetic field in the strained graphene. Here the overall out-of-plane deflection of the tent (or the graphene drumhead) is less significant. As the STM tip size keeps increasing beyond a certain value, the STM tip starts to interact with the graphene in a way similar to a nearly flat surface. As a result, further increases in STM tip size would not significantly change the interaction region between STM tip and membrane, which gives rise to the saturation trend as shown in Fig. 7(d).

On the other hand, for graphene drumheads with diameters larger than the threshold value, both the pseudomagnetic fields and the associated QD sizes become nearly independent of the graphene drumhead diameter, as is evident from the plateaus in the curves shown in Figs. 7(a)-(c). The plateau values of both the intensity of pseudomagnetic field and the associated QD size are higher for a larger STM probe tip diameter, although the increases are rather insignificant. This can be understood as follows: even though the amplitude of out-of-plane deflection of the graphene drumhead increases as the graphene size increases, the portion of graphene with high strain concentration occurs only in a rather localized region in the vicinity of the STM probe tip with a size and strain field profile that are nearly insensitive to the graphene size. To further demonstrate such an understanding, we employ the finite element method (FEM) to reveal the tip-induced strain distribution in the graphene drumhead. Equation 5 suggests that the pseudomagnetic field depends on the spatial distribution of strain components $u_{rr}$ and $u_{\theta\theta}$.

In the finite element method, the graphene drumhead is modeled with general purpose, large-strain quadrilateral shell elements (S4R), which allow for finite membrane strain [24]. The graphene drumhead has a thickness of 1.317 Å, a Young's modulus of 1 TPa and a Poisson's ratio of 0.19 [25]. The outer edge of the circular graphene drumhead is fixed. As the interaction force exerted by the STM tip on the graphene is quite complicated, we represent its effect by applying a uniformly distributed upward pressure at a central circular region (referred to as the 'loading zone') on the drumhead [Fig. 8(a)]. As suggested by the CG simulations, when the graphene size is much larger than the STM tip size, the size of the loading zone is assumed to be constant. In the finite element simulations, the diameter of the loading zone is set to be 10 nm, estimated from the critical contact area in the CG simulations with a STM tip of 110 nm in diameter. The amplitude of the applied pressure is estimated from the maximum of a vdW force between the CG tip bead and CG graphene beads. The finite element simulations are carried out using ABAQUS software (see disclaimer below).

Figure 8(b) plots the FEM simulation results of the contour of the two strain components $u_{rr}$ and $u_{\theta\theta}$ in circular graphene drumheads of three different diameters (500 nm, 900 nm and 1100 nm) which are greater than the threshold diameter of about 400 nm. By zooming in on a circular region of diameter of 20 nm at the center of the graphene (highlighted by red circles in Fig. 8(b)), it is shown that both the magnitude and spatial distribution of the two polar strain components $u_{rr}$ and $u_{\theta\theta}$ in such a circular region are nearly the same for the three graphene drumheads of different diameters, which explains the nearly constant pseudomagnetic field intensity as well as the QD size as the graphene drumhead is larger than a threshold diameter. On one hand, the FEM findings reveal that as the size of the graphene drumhead becomes large enough compared with the STM tip, the strain field in the graphene, within the same area beneath the tip, becomes less and less dependent on the drumhead size, therefore dictating the saturation trend of the QD size as a function of drumhead size. On the other hand, the agreement between CG and FEM results further suggests that continuum mechanics modeling indeed can provide reasonable prediction to the STM-tip-induced localized deformation in the graphene membrane, and thus offer reasonable insight on the resulting pseudomagnetic field in graphene. This feature becomes particularly attractive in practice for a graphene device with a size too large to be modeled even by the scalable CG, for which FEM can be an effective modeling approach to study its device behavior.

*IV.3. Effect of back-gate force and STM tip location on the strain-induced pseudomagnetic field*

Our experiments show that a back-gate voltage applied to the graphene device shown in Fig. 1(a) can tune the deformation of the graphene drumhead. For example, the back-gate voltage exerts a distributed force on the graphene drumhead that counteracts the vdW attractive force from the STM tip [Fig. 2]. As a result, the overall amplitude of the out-of-plane deflection of the graphene drumhead decreases as the back-gate voltage increases. The graphene drumhead can even dip down into the pit on the $SiO_2$ substrate if a sufficiently high back-gate voltage is applied. What remains unclear is if the back-gate voltage can also change the strain-induced pseudomagnetic field. To address this issue, we carry out CG simulations to model the deformation of the graphene drumhead subject to both the STM tip interaction and a back-gate voltage. The effect of the back-gate voltage is represented by a uniform force applied downward on each CG graphene bead.

Figure 9(a) compares the deflections of the graphene drumhead across its diameter in three cases: zero back-gate force as in Fig. 6(a), with a back-gate force of $9.65 \times 10^{-4}$ eV/ Å, and $1.61\ 10^{-3}$ eV/ Å per CG bead, respectively. It is evident that the back-gate voltage effectively

"pulls down" the graphene drumhead. The higher the back-gate voltage (i.e., larger back-gate force), the more the graphene drumhead dips downward. However, the shape of the cusp formed beneath the STM tip in all three cases appears rather similar, indicating a highly localized strain effect. Figures 9(b)-9(c) show the corresponding pseudomagnetic fields for the two cases with a back-gate force. It turns out that both the intensity and the spatial distribution (e.g., threefold clover-leaf symmetry) of the pseudomagnetic field, and therefore the corresponding QD size, remain nearly the same as in the case of no back-gate voltage [Fig. 6(e)]. This result suggests that a back-gate voltage only affects the overall deformed shape of the graphene drumhead but has nearly no effect on the STM tip-induced localized strain field and thus the pseudomagnetic field. Similar observations also hold for the cases where the STM tip is positioned in an off-centered location. Figure 9(d) shows that as the STM tip navigates away from the centered location on the graphene drumhead, the overall deflection of the graphene decreases and the global rotational symmetry of the deformed shape of the graphene disappears, but the highly localized deformation near the cusp beneath the STM tip remains nearly the same. Figures 9(e)-9(f) further reveal that the resulting pseudomagnetic field when the STM tip is off-center-positioned (even quite close to the fixed edge, e.g., 490 nm off-centered as shown in Fig. 9(f)) is almost the same as when the tip is center-positioned. The above results reveal that the STM tip-induced pseudomagnetic field and the corresponding QD are nearly independent of the STM tip location and the back-gate voltage.

## V. Concluding remarks

We conducted systematic coarse grained simulations to study the STM tip-induced deformation in a suspended graphene drumhead with a diameter up to 1.1 μm, which enables investigations of the strain-induced pseudomagnetic field and the associated spatially confined

QD generated in the graphene drumheads. The CG simulation results are validated by excellent agreement with both the results from the fully atomistic simulation of a scaled-down model and the STM/STS experimental measurements. A parametric study via comprehensive CG simulations reveals the dependence of the strain-induced pseudomagnetic field and the associated QD size on the diameters of the graphene drumhead and the STM probe tip. There exists a threshold size of graphene drumhead, below which the intensity of the pseudomagnetic field increases but the associated QD size decreases as the graphene drumhead size decreases, and above which both the pseudomagnetic field and the associated QD become nearly independent of the graphene drumhead size. There exists a threshold size of the STM probe tip, below which the intensity of the pseudomagnetic field increases but the associated QD size decreases as the STM probe tip size decreases, and above which both the pseudomagnetic field and the associated QD become nearly independent of the STM probe tip size. These results suggest two possible regimes of STM tip-induced spatially confining QDs in a graphene drumhead, in one of which a tunable QD size and intensity can be explored, and in another of which a robust spatially confining QD that can maneuver in association with the motion of the STM tip becomes possible. Results from the present study offer quantitative guidance for further explorations of strain engineering of suspended graphene drumheads.

**Acknowledgement:** TL, ZS and YH acknowledge the support by the National Science Foundation (Grant Numbers: 1069076 and 1129826). ZS thanks the support of the Clark School Future Faculty Program and Graduate Student Summer Research Fellowships at the University of Maryland. YH acknowledges the support of Dean's Fellowship from Clark School at the University of Maryland. NK acknowledges support under the Cooperative Research Agreement between the University of Maryland and the National Institute of Standards and Technology

**Disclaimer:** Certain commercial equipment, instruments, materials or computational tools are identified in this paper in order to specify the experimental or simulation procedures adequately. Such identification is not intended to imply recommendation or endorsement by the National Institute of Standards and Technology, nor is it intended to imply that the materials or equipment identified are necessarily the best available for the purpose.

FIG. 1. (a) Optical image of the back-gated graphene device. The device consists of a single graphene layer placed over an array of pits (1.1 µm in diameter, 100 nm in depth) etched in a $SiO_2$/Si substrate. (b) Scanning electron microscope image of a single-layer graphene flake placed over the array of pits. (c) Schematic of the deformation of the suspended graphene drumhead induced by the STM probe tip (not to scale).

FIG. 2. STM images of the deformed graphene membrane induced by the competing forces from the probe tip pulling up on the membrane and the back gate electric field pulling down on the membrane. The STM images are shown for various gate potentials from 0 V to 50 V. At 50 V the membrane is being pulled into the pit in the substrate.

FIG. 3. (a) STM topographic profiles obtained while scanning from graphene supported onto the SiO2 substrate to the suspended region over a pit in the substrate as a function of tunneling bias at a fixed gate potential of 0 V and tunneling current of 50 pA. (b) SEM image of the STM probe tip showing a probe diameter of ≈ 110 nm.

FIG. 4. STS $dI/dV$ gate maps of graphene measured at 8 T magnetic field for (a) graphene-on-$SiO_2$ (≈ 50 nm from the pit's edge), (b) graphene-on-$SiO_2$ (≈ 10 nm from the pit's edge), (c) suspended graphene (≈ 10 nm from the pit's edge into the membrane) [2]. While Landau quantization is observed on supported graphene in (a), multiple quartets of single electron charging peaks characteristic to Coulomb blockade physics are observed on suspended graphene in (c). In the nearest proximity to the pit's edge graphene can be delaminated from the $SiO_2$ surface at low gate voltages when the vdW upward pulling force dominates the downward electrostatic force from the back gate. Within the range of -20 V < $V_{GATE}$ < 20 V the graphene is pulled up by 4 nm to 5 nm from the $SiO_2$ surface and a signature of QD formation is observed in the spectra in (b). (d) $dI/dV$ versus $V_B$ spectra at $V_{GATE}$ = 5 V from the gate map in (c) showing multiple four-fold groups of charging resonances. (e) Measured peak positions from the spectra in (d). Error bars are smaller than data symbols. (f) The quantum dot addition energies corresponding to the difference in $dI/dV$ peak positions in (e). Energies are converted from bias

voltages using the lever arm, $E = \alpha V_B$, where $\alpha = 0.45 \pm 0.03$ [2].

FIG. 5. (a) Carbon atom representation of different orders of CG graphene beads. Purple triangles serve as the visual guide showing the inherent correlation between different orders of CG beads. (b) Coarse grained scheme lattices composed of different orders of CG graphene beads, which preserve the hexagon pattern. Beads are colored in accordance with (a). (c) Left: CG computational model. The STM tip is coarse-grained as a single spherical bead and graphene is composed of CG beads of a given CG order as depicted in (a). Right: Comparison of the calculated deflection of the graphene membrane of 50 nm in diameter between the CG method and the fully atomistic molecular dynamics simulations in Ref. [2].

FIG. 6. Simulation results of STM tip-induced deformation in graphene drumheads and the resulting pseudomagnetic field. Here there is no electrostatic force from a back-gate electrode and the STM probe tip is centered. (a) Out of plane deflection contour. (b-d): Strain components $u_{xx}$, $u_{yy}$ and $u_{xy}$, respectively. (e) Top view of the associated pseudomagnetic field. (f) Zoomed-in perspective view of the pseudomagnetic field in the center region of the graphene drumhead.

FIG. 7. (a)-(c) Intensity of STM tip-induced pseudomagnetic field (blue diamonds) and the corresponding QD size (red dots) as a function of the diameter of the graphene drumhead, for various STM probe tip diameters. The dashed line in (a), (b) and (c) estimates a threshold size of the graphene drumhead that delineates two regimes of the dependence of pseudomagnetic field on graphene drumhead diameter. (d) The dependence of the intensity of STM tip-induced pseudomagnetic field (blue diamonds) and the corresponding QD size (red dots) on the diameter of the STM probe tip, for a graphene drumhead of diameter of 1100 nm. The dashed arrows in (d) outline the saturation trend of such a dependence, as the size of the STM probe tip increases. Here there is no electrostatic force from a back-gate electrode and the STM probe tip is positioned above the center of the graphene drumhead.

FIG. 8. Finite element simulations reveal the highly localized strain distribution in a circular region in the vicinity of the STM probe tip. (a) Schematics of the finite element simulation model. The red circle denotes the fixed boundary and the smaller circle near the center defines the loading zone, where the effect of tip/graphene interaction is represented by a uniformly distributed pressure. (b) Strain component contours in a cylindrical coordinate system for graphene drumheads of three different diameters. For visual clarity, the strain distribution inside the loading zone (white circle) is not shown. All zoomed-in contours have the same size, which reveal the rather similar strain field in the vicinity of the STM tip, which is the origin of the insensitivity of the pseudomagnetic field to the graphene drumhead size, when the graphene is much larger than the STM probe tip.

FIG. 9. CG simulation results of the STM tip-induced deformation of the graphene drumhead and associated pseudomagnetic field when a back-gate force is applied and the STM tip is at off-centered locations. Here the diameter of the graphene drumhead is 1.1 μm. (a) and (d) show the deflection of the graphene drumhead across its diameter. (b), (c), (e) and (f) show the corresponding pseudomagnetic fields. Case #0: center-positioned STM probe tip with zero back-

gate force, replotted from Fig. 6 for comparison. Case #1 and case #2: center-positioned STM probe tip with downward back-gate forces of $9.65\times10^{-4}$ eV/Å and $1.61\times10^{-3}$ eV/Å per CG bead, respectively. Case #3 and case #4: tip off-center-positioned by 440 nm and 490 nm with zero back-gate force, respectively. Insets show zoomed-in images of the pseudomagnetic field in the vicinity of the STM probe tip.

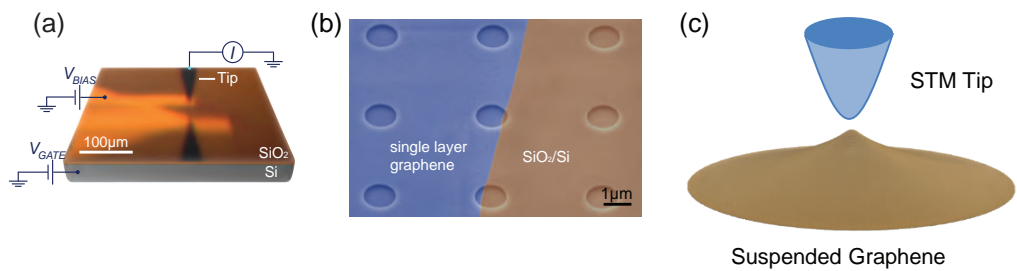

FIG. 1. (a) Optical image of the back-gated graphene device. The device consists of a single graphene layer placed over an array of pits (1.1 μm in diameter, 100 nm in depth) etched in a SiO$_2$/Si substrate. (b) Scanning electron microscope image of a single-layer graphene flake placed over the array of pits. (c) Schematic of the deformation of the suspended graphene drumhead induced by the STM probe tip (not to scale).

*print full 1.5 columns

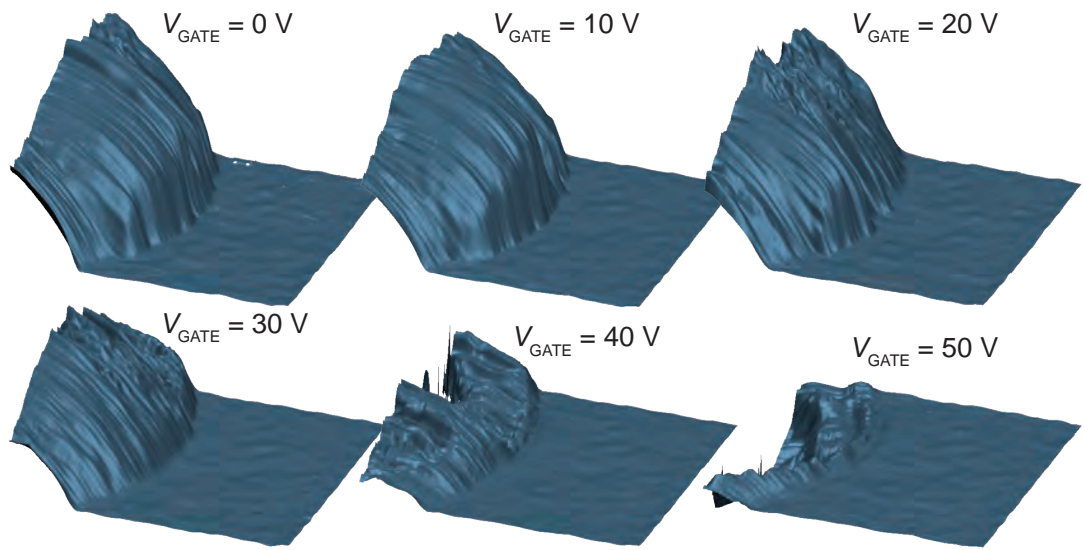

Fig. 2. STM images of the deformed graphene membrane induced by the competing forces from the probe tip pulling up on the membrane and the back gate electric field pulling down on the membrane. The STM images are shown for various gate potentials from 0 V to 50 V. At 50 V the membrane is being pulled into the pit in the substrate.

*print full double column

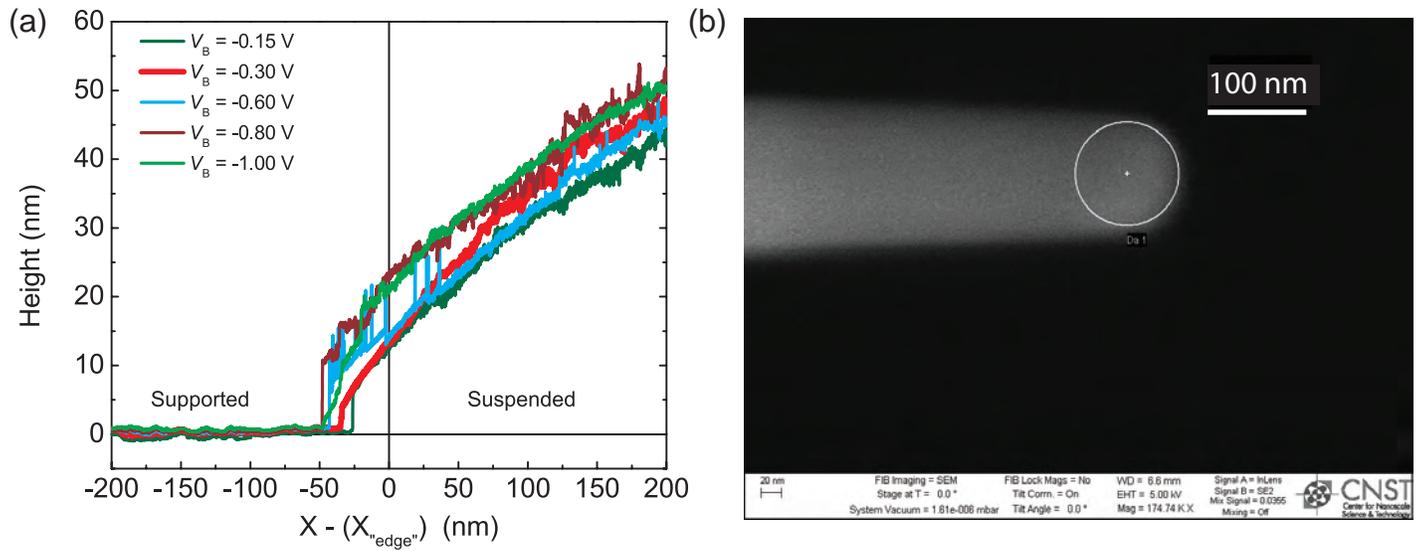

Fig. 3. (a) STM topographic profiles obtained while scanning from graphene supported onto the SiO$_2$ substrate to the suspended region over a pit in the substrate as a function of tunneling bias at a fixed gate potential of 0 V and tunneling current of 50 pA. (b) SEM image of the STM probe tip showing a probe diameter of ≈ 110 nm.

*print full double column

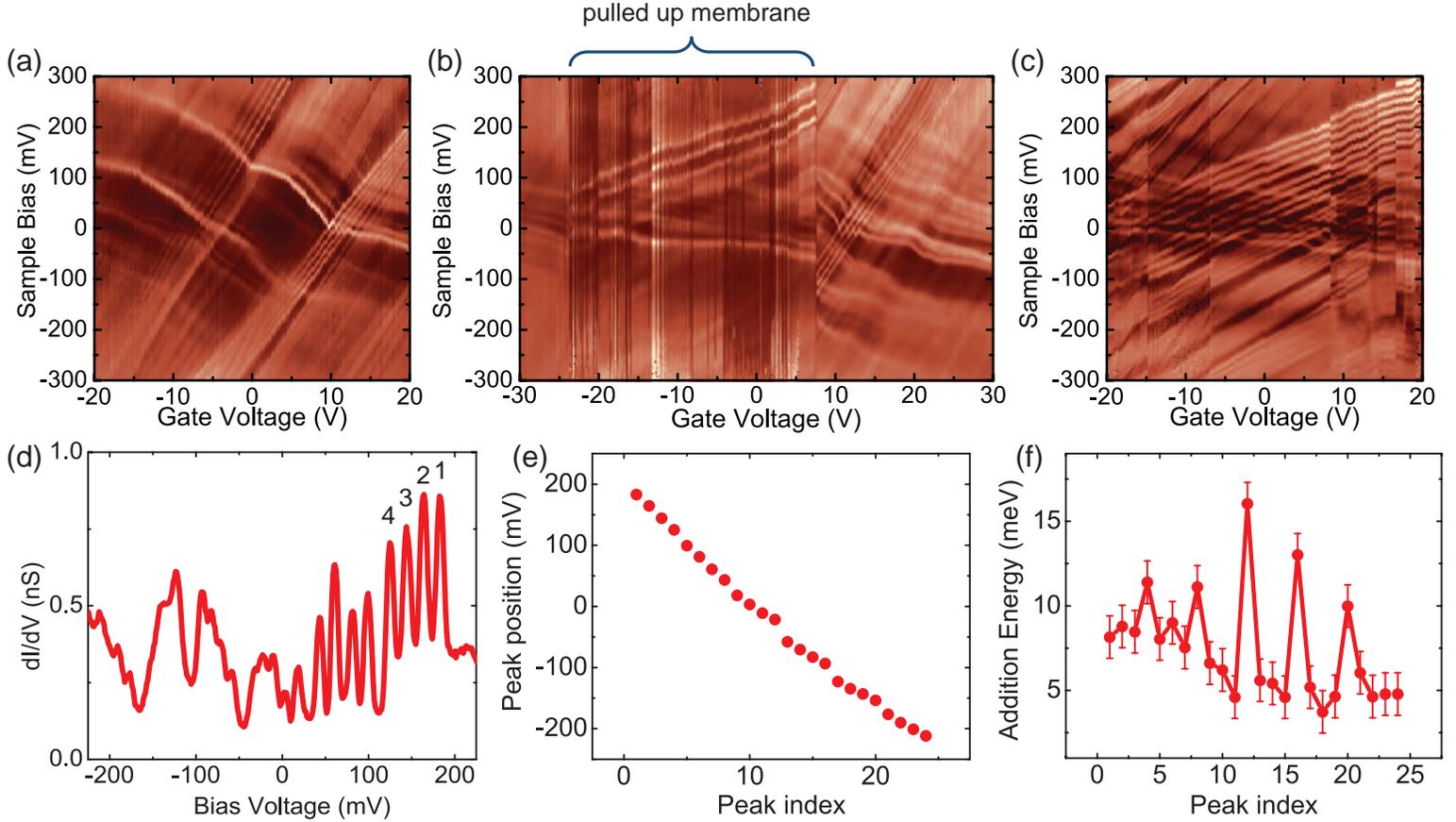

FIG. 4. STS *dI/dV* gate maps of graphene measured at 8T magnetic field for (a) graphene-on-SiO$_2$ (50 nm from the pit's edge), (b) graphene-on-SiO$_2$ (10 nm from the pit's edge), (c) suspended graphene (10nm from the pit's edge into the membrane) [2]. While Landau quantization is observed on supported graphene in (a), multiple quartets of single electron charging peaks characteristic to Coulomb blockade physics are observed on suspended graphene in (c). In the nearest proximity to the pit's edge graphene can be delaminated from the SiO$_2$ surface at low gate voltages when the vdW upward pulling force dominates the downward electrostatic force from the back gate. Within the range of -20 V < $V_{GATE}$ < 20 V the graphene is pulled up by 4 nm to 5 nm from the SiO$_2$ surface and a signature of QD formation is observed in the spectra in (b). (d) *dI/dV* versus $V_B$ spectra at $V_{GATE}$ = 5 V from the gate map in (c) showing multiple four-fold groups of charging resonances. (e) Measured peak positions from the spectra in (d). Error bars are smaller than data symbols. (f) The quantum dot addition energies corresponding to the difference in *dI/dV* peak positions in (e). Energies are converted from bias voltages using the lever arm, $E= \alpha V_B$, where $\alpha=0.45\pm0.03$ [2].

*print full double column

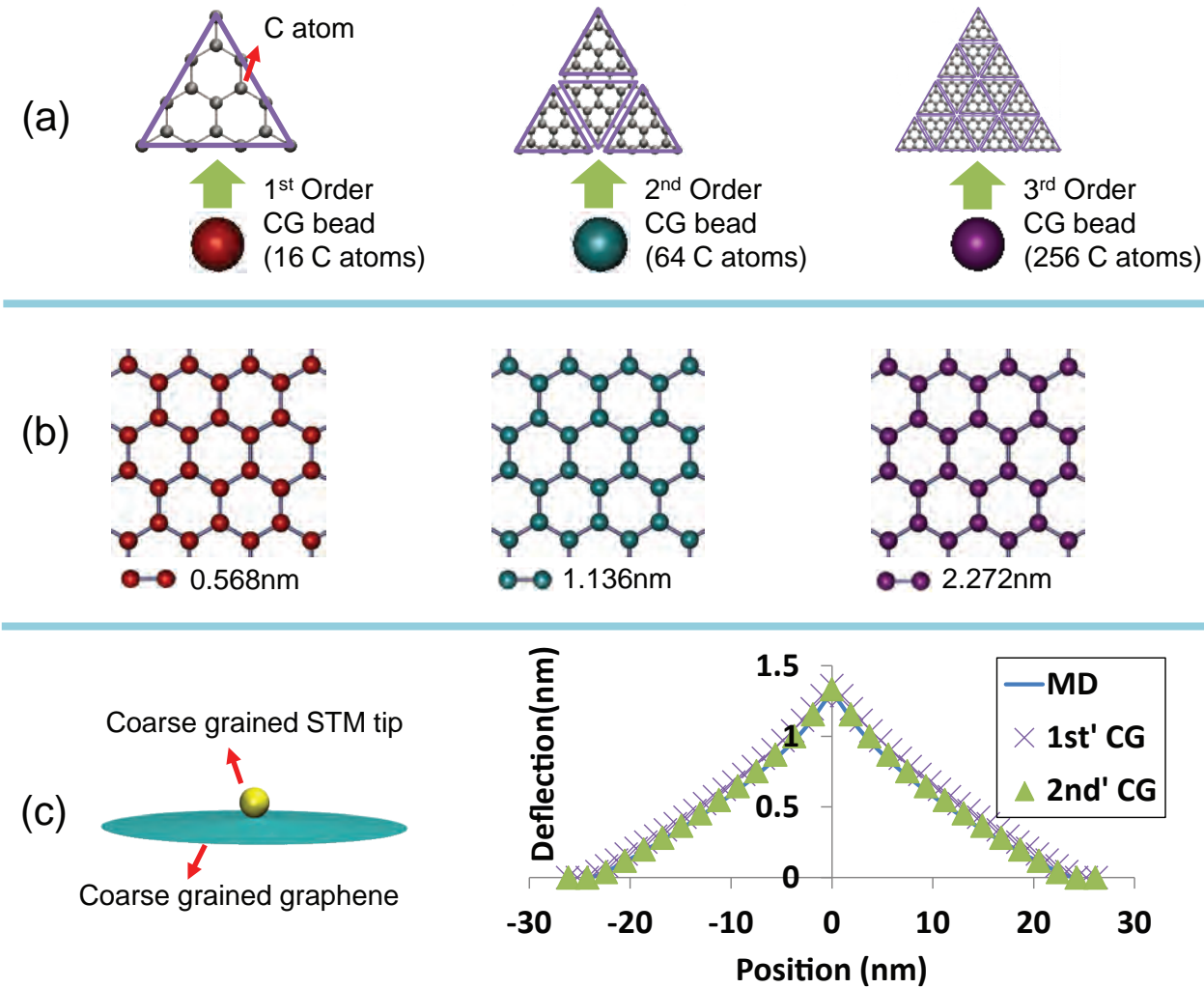

FIG. 5. (a) Carbon atom representation of different orders of CG graphene beads. Purple triangles serve as the visual guide showing the inherent correlation between different orders of CG beads. (b) Coarse grained scheme lattices composed of different orders of CG graphene beads, which preserve the hexagon pattern. Beads are colored in accordance with (a). (c) Left: CG computational model. The STM tip is coarse-grained as a single spherical bead and graphene is composed of CG beads of a given CG order as depicted in (a). Right: Comparison of the calculated deflection of the graphene membrane of 50 nm in diameter between the CG method and the fully atomistic molecular dynamics simulations in Ref. [2].

*print full 2 columns

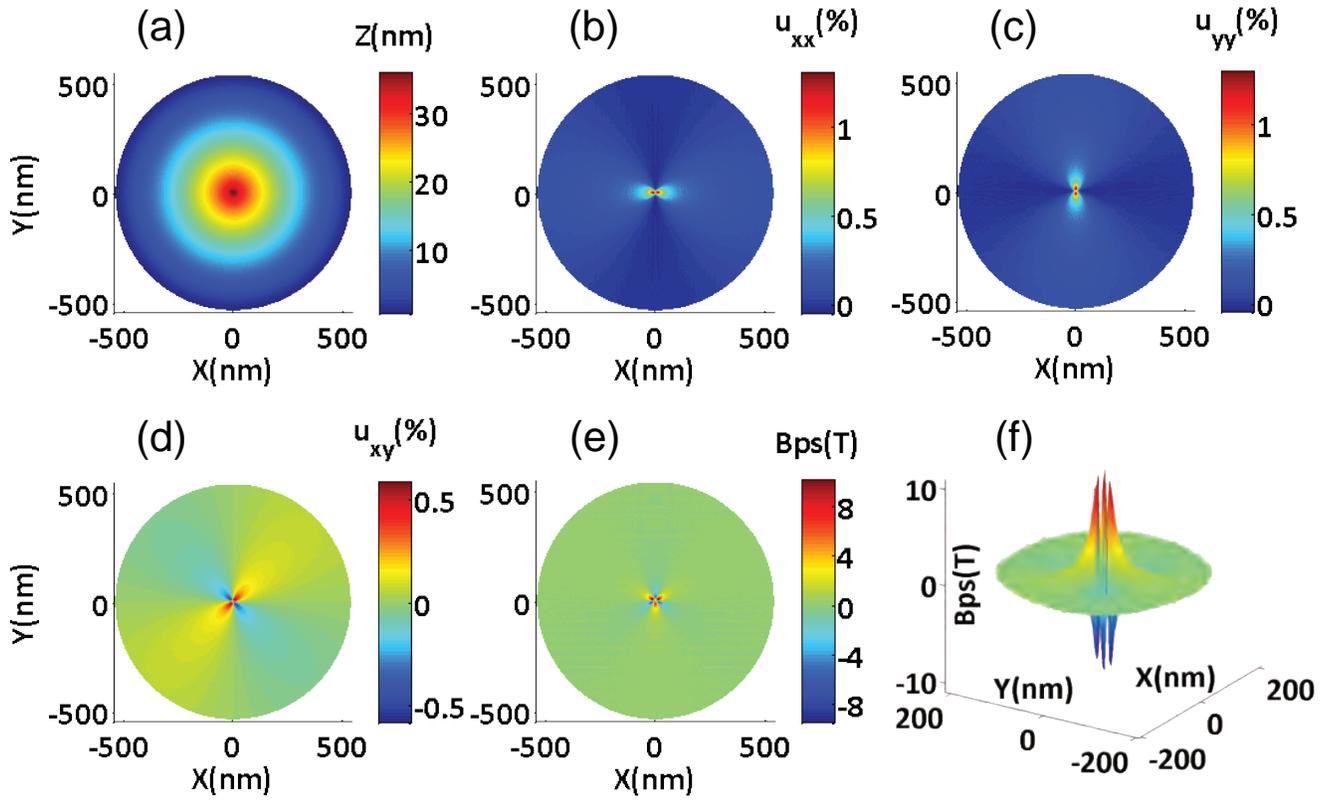

FIG. 6. Simulation results of STM tip-induced deformation in graphene drumheads and the resulting pseudomagnetic field. Here there is no electrostatic force from a back-gate electrode and the STM probe tip is centered. (a) Out of plane deflection contour. (b-d): Strain components $u_{xx}$, $u_{yy}$ and $u_{xy}$, respectively. (e) Top view of the associated pseudomagnetic field. (f) Zoomed-in perspective view of the pseudomagnetic field in the center region of the graphene drumhead.

*print full 2 columns

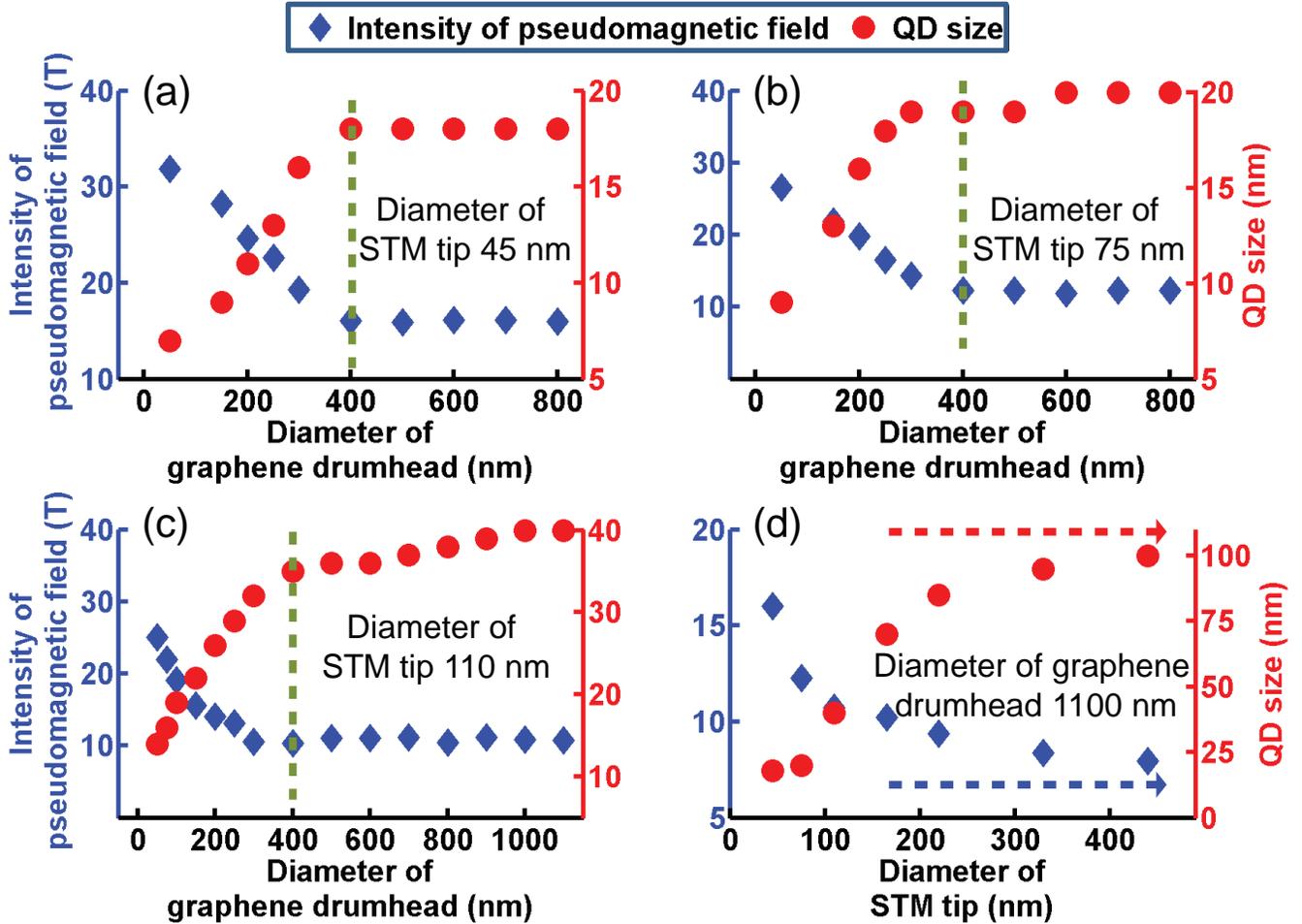

FIG. 7. (a)-(c) Intensity of STM tip-induced pseudomagnetic field (blue diamonds) and the corresponding QD size (red dots) as a function of the diameter of the graphene drumhead, for various STM probe tip diameters. The dashed line in (a), (b) and (c) estimates a threshold size of the graphene drumhead that delineates two regimes of the dependence of pseudomagnetic field on graphene drumhead diameter. (d) The dependence of the intensity of STM tip-induced pseudomagnetic field (blue diamonds) and the corresponding QD size (red dots) on the diameter of the STM probe tip, for a graphene drumhead of diameter of 1100 nm. The dashed arrows in (d) outline the saturation trend of such a dependence, as the size of the STM probe tip increases. Here there is no electrostatic force from a back-gate electrode and the STM probe tip is positioned above the center of the graphene drumhead.

*print full 1.5 columns

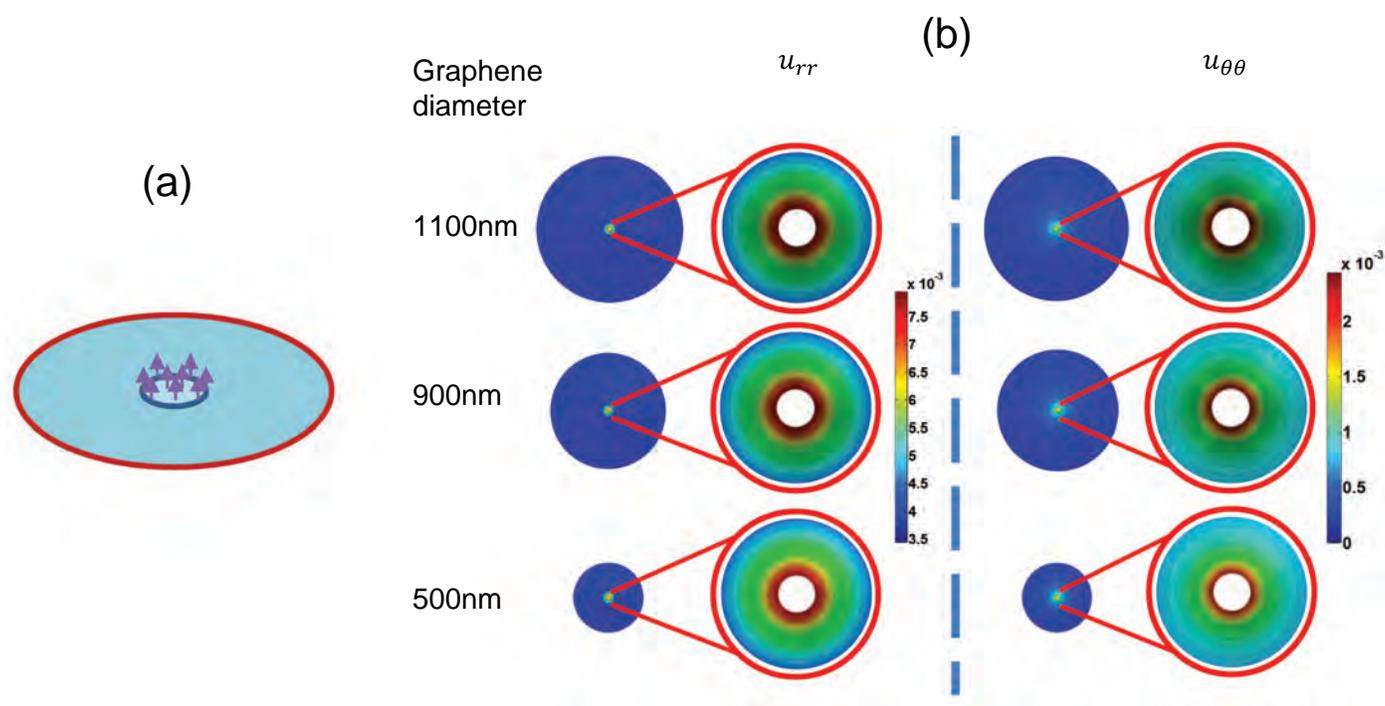

FIG. 8. Finite element simulations reveal the highly localized strain distribution in a circular region in the vicinity of the STM probe tip. (a) Schematics of the finite element simulation model. The red circle denotes the fixed boundary and the smaller circle near the center defines the loading zone, where the effect of tip/graphene interaction is represented by a uniformly distributed pressure. (b) Strain component contours in a cylindrical coordinate system for graphene drumheads of three different diameters. For visual clarity, the strain distribution inside the loading zone (white circle) is not shown. All zoomed-in contours have the same size, which reveal the rather similar strain field in the vicinity of the STM tip, which is the origin of the insensitivity of the pseudomagnetic field to the graphene drumhead size, when the graphene is much larger than the STM probe tip.

*print full 2 columns

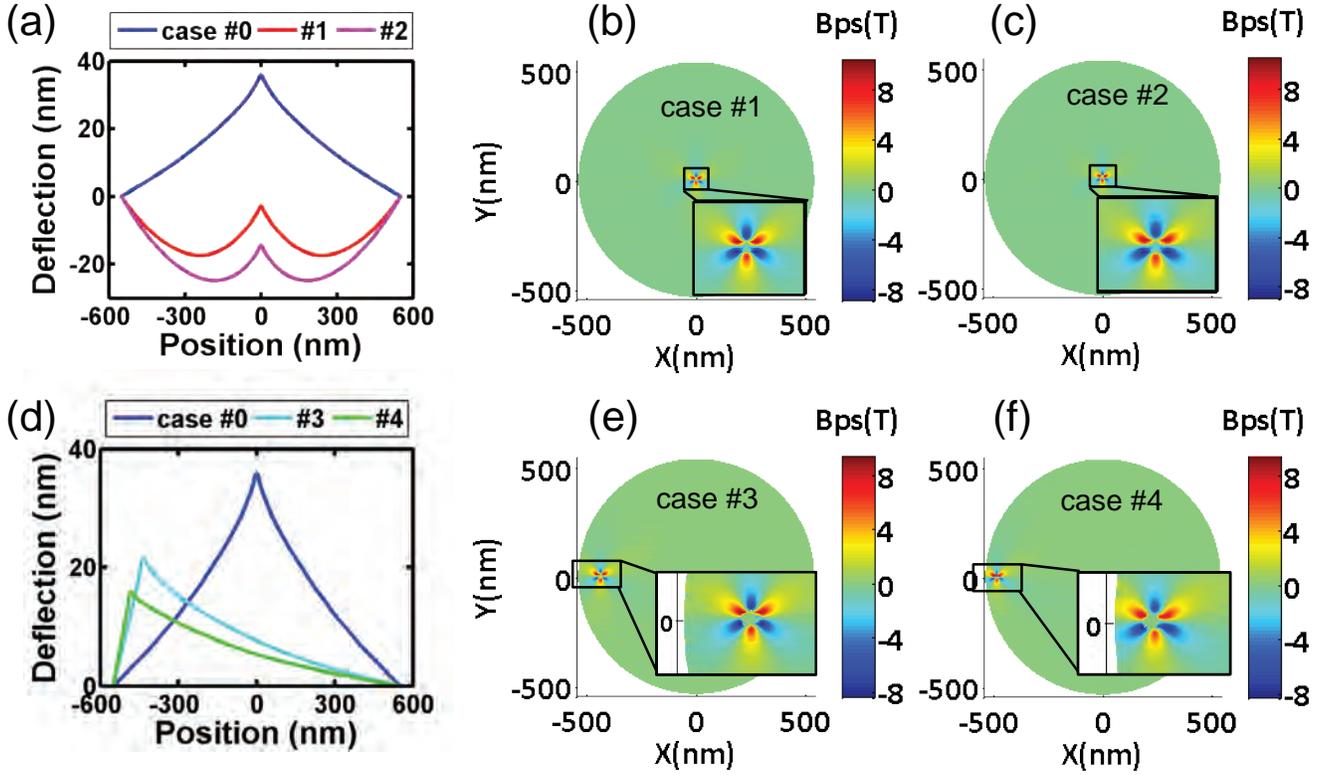

FIG. 9. CG simulation results of the STM tip-induced deformation of the graphene drumhead and associated pseudomagnetic field when a back-gate force is applied and the STM tip is at off-centered locations. Here the diameter of the graphene drumhead is 1.1 μm. (a) and (d) show the deflection of the graphene drumhead across its diameter. (b), (c), (e) and (f) show the corresponding pseudomagnetic fields. Case #0: center-positioned STM probe tip with zero back-gate force, replotted from Fig. 6 for comparison. Case #1 and case #2: center-positioned STM probe tip with downward back-gate forces of $9.65 \times 10^{-4}$ eV/Å and $1.61 \times 10^{-3}$ eV/Å per CG bead, respectively. Case #3 and case #4: tip off-center-positioned by 440 nm and 490 nm with zero back-gate force, respectively. Insets show zoomed-in images of the pseudomagnetic field in the vicinity of the STM probe tip.

\*print full 2 columns